# Reversible Image Data Hiding Using Lifting Wavelet Transform and Histogram Shifting

S. Kurshid Jinna
Professor, Dept of Computer Science & Engineering
PET Engineering College
Vallioor, Tirunelveli, India
kurshidjinna@gmail.com

Dr. L. Ganesan
Professor, Dept of Computer Science & Engineering
A.C College of Engineering & Technology
Karaikudi, India
csehod@gmail.com

*Abstract*- **A method of lossless data hiding in images using integer wavelet transform and histogram shifting for gray scale images is proposed. The method shifts part of the histogram, to create space for embedding the watermark information bits. The method embeds watermark while maintaining the visual quality well. The method is completely reversible. The original image and the watermark data can be recovered without any loss.**

*Keywords: Data Hiding, Histogram shifting, reversible watermarking, Integer-integer wavelet transforms.*

## I. Introduction

The reversible watermarking algorithms are developed from the time it was suggested by its pioneers. Fridrich et al, Jun Tian and Ni et al are pioneers in the field.

Ni et al. [1] proposed an image lossless data hiding algorithm using pairs of zero-points and peak-points, in which the part of an image histogram is shifted to embed data. lossless data embedding algorithm based on the histogram shifting in spatial domain is proposed. J.Fridrich and M. Goljan suggested general methodologies for lossless embedding that can be applied to images as well as any other digital objects. The concept of lossless data embedding can be used as a powerful tool to achieve a variety of necessary tasks, including lossless authentication using fragile watermarks[2].

J. Tian calculates the differences of neighboring pixel values, and selects some difference values for the difference expansion (DE) for reversible data embedding as suitable pairs for data embedding. Pairs which do not affect the algorithm for lossless embedding and extraction are used and is indicated with the help of location map[3].

Xuan et al.[4] proposed the lossless embedding using the integer wavelet transform (IWT) and histogram medication using a threshold point for embedding limit. G. Xuan and Y. Q. Shi proposed a histogram shifting method for image lossless data hiding in integer wavelet transform domain. This algorithm hides data into wavelet coefficients of high frequency subbands. It shifts part of the histogram of high frequency wavelet subbands and embeds data by using the created histogram zero-point [5]. Chrysochos et al's scheme of reversible watermarking presents a method resistant to geometrical attacks [6].

Fallahpour M, Sedaaghi M proposes relocation of zeroes and peaks of the histogram of the image blocks of the original image to embed data in the spatial domain. Image is divided into varying number of blocks as required and the performance is analysed. [7]

Xianting Zenga,Lingdi Ping and Zhuo Li proposed scheme based on the difference histogram shifting to make space for data hiding. Differences of adjacent pixels are calculated by using different scan paths. Due to the fact that the grayscale values of adjacent pixels are close to each other, the various-directional adjacent pixel difference histogram contains a large number of points with equal values, data hiding space is obtained[8].

## II Integer-To-Integer Wavelet Transforms

In conventional wavelet transform reversibility is not achieved due to the floating point wavelet coefficients we get after transformation. When we take the inverse transform the original pixel values will get altered.

When we transform an image block consisting of integer-valued pixels into wavelet domain using a floating-point wavelet transform and the values of the wavelet coefficients are changed during watermark embedding, the corresponding watermarked image block will not have integer values. When we truncate the floating point values of the pixels, it may result in loss of information and reversibility is lost. The original image cannot be reconstructed from the watermarked image.

In conventional wavelet transform done as a floating-point transform followed by a truncation or rounding it is impossible to represent transform coefficients accurately. Information will be potentially lost through forward and inverse transforms.

In view of the above problems, an invertible integer-to-integer wavelet transform based on lifting is used in the proposed scheme. It maps integers to integers which are preserved in both forward and reverse transforms. There is no loss of information. Wavelet or subband decomposition associated with finite length filters is obtained by a finite number of primal and dual lifting followed by scaling.

### III Wavelet Histogram Shifting

Integer Wavelet transforms of the original image is taken. In the subband wavelet histogram data is to be embedded. In the histogram the horizontal axis(X) represents the wavelet coefficients value and the vertical axis(Y) represents the number of occurrence of each coefficient's value. The wavelet histogram normally exhibits a Laplacian distribution nature with a peak point and sloping on either side. Peak in wavelet histogram is usually at coefficient value '0'

Embedding can be done on both the sides of the histogram to get the required embedding capacity.

Data embedding is done by modifying some of the coefficient values of the wavelet domain to it's neighboring value by shifting a portion of the histogram. This gives a good visual quality and thereby a better PSNR between original image and watermarked image.

To embed data we choose the peak point of the histogram and call it as P. Figure 1 shows a vacant point is created at Peak+1.This is done by shifting all points with value Peak+1 and above one position to the right. Now all the IWT coefficients are scanned and whenever a coefficient with value peak is encountered, '0' is embedded by leaving it as such and '1' is embedded by changing it's value to peak+1.This is repeated till all the points with value Peak are over. Then a new peak is created by shifting to the right and data is embedded as per the algorithm. We choose the peak point so that payload is maximized.

All the high frequency wavelet subbands can be utilized to get maximum capacity. The same process can be done on the left side of the histogram Peak to embed more watermark bits. A reverse algorithm is applied for extracting the watermark data.

After water mark bits are extracted, shifting is done towards the left each time after data extraction so that the original coefficient values are restored. This guarantees complete reversibility and the original image can be exactly reconstructed without loss.

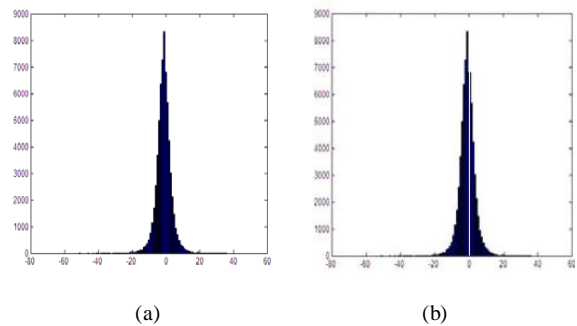

(a)  (b)

Figure 1 Illustration of wavelet Histogram, (a) Maximum point is at Peak, (b) Histogram with zero point created at peak +1

### IV. Proposed Method
#### A. Embedding Method

For the wavelet transformed image sub bands histogram is taken. Now we can start embedding using the following steps. For the selected sub band, set P = Peak of the histogram coefficients.

Create a zero point at P+1 so that no point in the histogram has the value P+1.To create the zero point shift all coefficients with value P+1 and above to one position right. This makes P+1 as P+2, and the original P+2 to P+3 and so on.

1. Now positions P and P+1 are chosen to embed data.
2. Read the n watermark bits $W_b$ where $0 < b < n-1$.
3. Check $W_b = 0$, then '0' is embedded in the coefficient with value P by leaving it unchanged as P.
4. Check $W_b = 1$, then '1' is embedded in the coefficient with value P by changing it to value P+1.
5. Point P+1 gets slowly filled up depending upon the number of $W_b$ bits with value 1.
6. Go to histogram of the other sub bands to be marked and repeat the same process.

7. While to-be-embedded watermark bits are still remaining, set P=Peak+2 and go to step1. Otherwise stop.

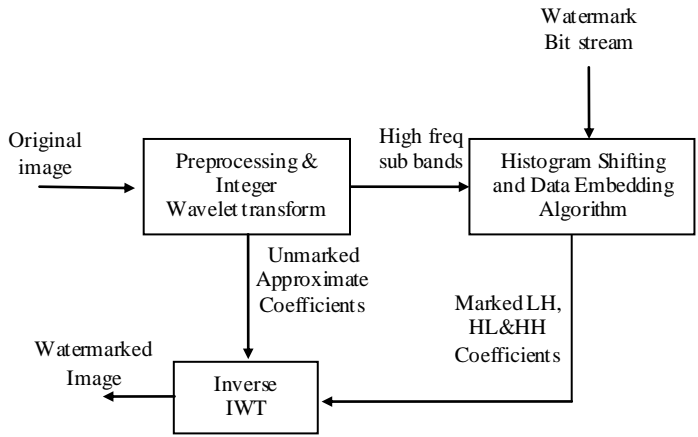

Figure 2 Embedding Method

Figure 2 Shows the original image is decomposed into it's sub bands using integer wavelet transform After preprocessing IWT is used to ensure complete reversibility. The high frequency sub bands (horizontal, Vertical and Diagonal) are used for data embedding. Each sub band is used one after the other to meet the required embedding capacity. Watermark bits that forms the payload is embedded into these sub bands using the embedding algorithm. The low frequency unmarked approximate coefficients are then used along with the marked sub bands and Inverse IWT is taken to get the watermarked image.

*B. Extraction Method*

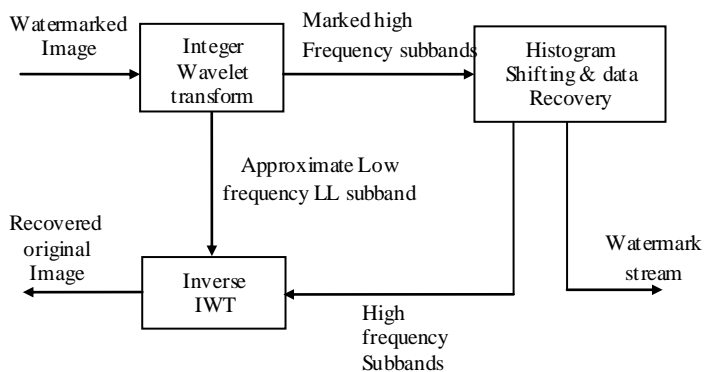

Figure 3 Extraction Method

The extraction method is shown in figure 3. Data extraction is the reverse process. Integer wavelet Transform is taken for the watermarked image. The watermarked high frequency sub bands are separated and using the Data extraction algorithm, the watermark bits are retrieved and the original sub bands are obtained. This is combined with the unmarked low frequency sub band to get the original image. This method is completely blind and reversible. Original image and the watermark data bits are obtained without any loss.

After wavelet decomposition of the watermarked image, histograms of the marked sub bands are taken. For the selected sub band, set Peak= Peak of the histogram coefficients.

1. P=Peak. Read the coefficients with value P and P+1. Whenever a coefficient with value P is read, extract watermark bit as $W_b = 0$ and leave P unaltered. Whenever a coefficient with value P+1 is read, extract watermark bit as $W_b = 1$ and change P+1 to P.
2. Shift all the coefficients with value P+2 and above one position to the left.
3. Go to histogram of the other marked sub bands and repeat the same process.
4. Set P = Peak+1.
5. While all watermark bits $W_n$ are not extracted go to step1. Otherwise stop.

V. Experimental Results and Discussion

Experiments are conducted using different 512 X 512 gray scale images and different wavelets.

Figure 4 shows image quality tested on different gray scale images after embedding around 1,00,000 bits.

Table 1 shows that cameraman image has a better embedding capacity than other images in the experiment. It also shows it has a better visual quality as far as Peak signal to noise ratio is concerned.

Figure 5 shows image quality tested on cameraman gray scale image after embedding different payload bits.

Figure 6 shows the image quality tested for different images using integer wavelet transform for different payloads using cdf2.2 wavelet. The sailboat image though has higher quality for the same payload compared to Lena image using lower payload, the image quality quickly falls down as payload is increased.

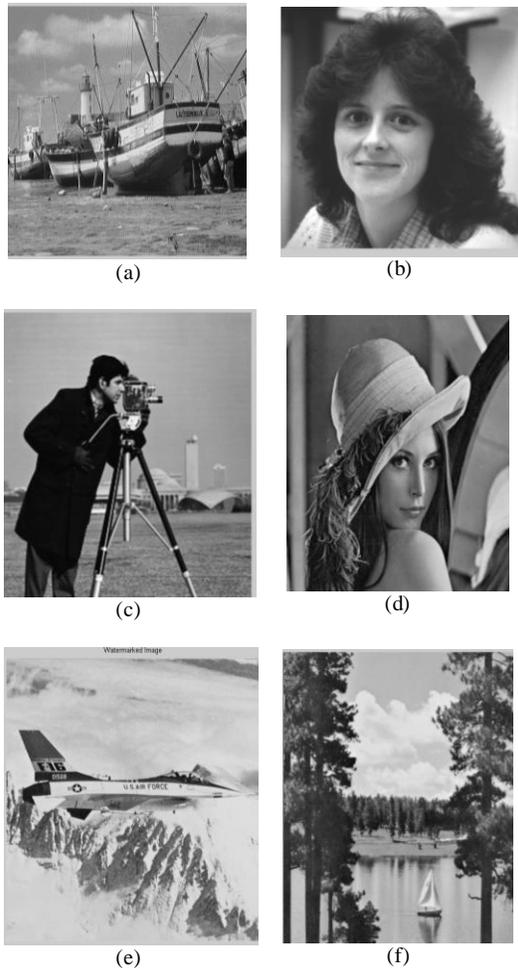

Figure 4 Watermarked Image with payload (bpp) 0.4 (a) Sail Boat PSNR 37.05 dB, (b) Woman Dark Hair PSNR 45.03 dB, (c) Camera Man PSNR 48.65 dB, (d) Lena PSNR 37.89 dB, (e) Jet Plane PSNR 42.35 dB, (f) Lake PSNR 36.42 dB

Table 1 Image Quality Tested for Different Grayscale Images for each payload using Cdf2.2 wavelet

| Payload (bpp) | Lena PSNR (dB) | Cameraman PSNR (dB) | Woman Dark Hair. PSNR (dB) | Sail Boat PSNR (dB) |
|---|---|---|---|---|
| 0.1 | 46.3338 | 48.8519 | 47.5824 | 47.2015 |
| 0.15 | 45.9927 | 48.7678 | 47.5095 | 44.0572 |
| 0.2 | 43.6043 | 48.7591 | 47.5146 | 43.1822 |
| 0.25 | 42.1267 | 48.7421 | 47.4691 | 41.5522 |
| 0.3 | 40.1725 | 48.7014 | 47.4321 | 39.5865 |
| 0.35 | 39.3304 | 48.6990 | 45.7253 | 38.4327 |
| 0.4 | 37.8965 | 48.6552 | 45.0346 | 37.0477 |
| 0.45 | xxx | 48.5825 | 43.8402 | xxx |
| 0.5 | xxx | 46.5234 | 42.2365 | xxx |
| 0.55 | xxx | 42.1261 | xxx | xxx |

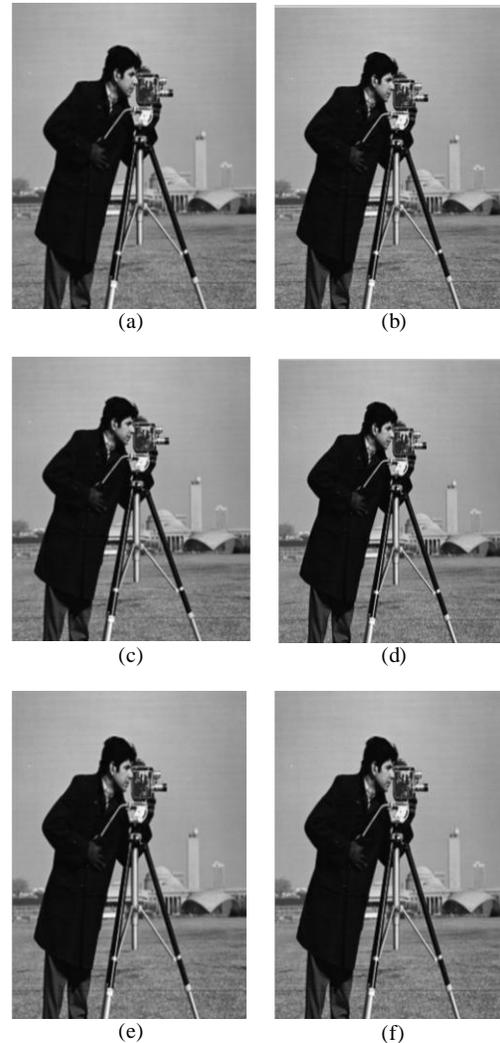

Figure 5 Watermarked Images of Cameraman (a) Original Image (b) Payload(bpp)0.2, PSNR 48.75dB, (c) Payload(bpp)0.3, PSNR 48.70 dB, (d) Payload(bpp)0.4, PSNR 48.65 dB, (e) Payload(bpp)0.5, PSNR 46.52 dB, (f) Payload(bpp)0.6, PSNR 42.13dB

Table II shows image quality tested for different payloads on the same image using different wavelets. Cdf2.2 performs better than other wavelets for the same payload. Image quality quickly changes when different wavelets are used. Performance in embedding measured using peak signal to noise ratio shows that bior6.8 has the minimum quality. The embedding capacity also varies when using different wavelets using different wavelets when the image is decomposed using db2 embedding stops in about 50,000 bits whereas cdf2.2 continues to embed over one lakhs bits. The same is illustrated in the graph.

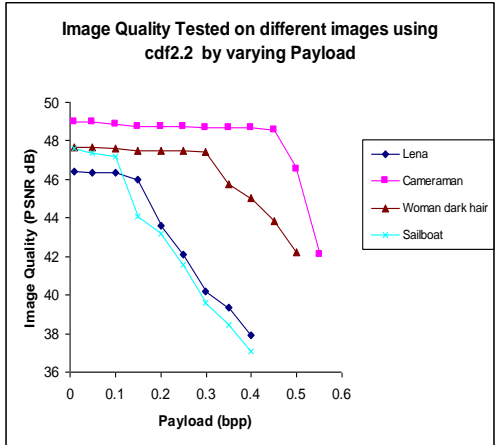

Figure 6 Image Quality Tested on Different Images Using cdf 2.2 by Various Payload (bpp)

Table II Comparison of performance of various wavelet families on Cameramen for different payload size

| Payload Bits | cdf2.2 PSNR (dB) | db2 PSNR (dB) | sym3 PSNR (dB) | bior3.3 PSNR (dB) | bior6.8 PSNR (dB) | rbio3.3 PSNR (dB) |
|---|---|---|---|---|---|---|
| 10000 | 48.96 | 44.63 | 43.05 | 46.85 | 39.93 | 43.50 |
| 15129 | 48.95 | 44.62 | 43.04 | 46.81 | 39.93 | 43.49 |
| 20164 | 48.91 | 44.61 | 43.03 | 46.78 | 39.93 | 43.48 |
| 25281 | 48.86 | 44.60 | 43.02 | 46.76 | 39.99 | 43.47 |
| 50176 | 48.75 | 44.56 | 42.93 | 46.61 | 39.94 | 43.46 |
| 75076 | 48.78 | xxx | 42.90 | 46.55 | 39.86 | 43.36 |
| 100489 | 48.67 | xxx | 42.83 | xxx | xxx | xxx |

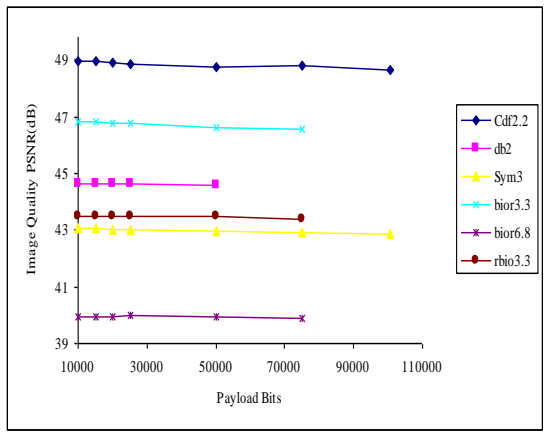

Figure 7 Image Quality Tested on Different Wavelets

Experiments were conducted on various 512x512 grayscale images to study the performance of various wavelets on the embedding algorithm. For a fixed payload of 25,000 bits embedded and tested, db1 performs best as shown in the table III. Bior6.8 has the minimum quality. A variation of about 10db in Peak signal to noise ratio exists while changing the wavelet family used for decomposing the original image for embedding. Also PSNR between the original image and the watermarked image varies depending on the image when using the same wavelet for decomposition.

Table III Image Quality Tested for Different Grayscale Images for fixed payload of 25000 bits

| Wavelets | Lena PSNR (dB) | Cameraman PSNR (dB) | Woman Dark Hair. PSNR (dB) | Sail Boat PSNR (dB) |
|---|---|---|---|---|
| db1 | 47.56 | 48.32 | 49.37 | 48.85 |
| cdf2.2 | 46.34 | 46.58 | 48.87 | 47.59 |
| bior3.3 | 45.37 | 45.53 | 46.75 | 46.38 |
| sym2 | 44.62 | 44.98 | 44.60 | 44.83 |
| db3 | 42.35 | 42.51 | 42.14 | 42.44 |
| sym3 | 41.12 | 41.16 | 43.02 | 42.51 |
| rbio3.3 | 41.09 | 41.32 | 43.51 | 42.87 |
| rbio6.8 | 40.45 | 40.79 | 40.48 | 40.66 |
| bior6.8 | 39.88 | 40.40 | 39.99 | 40.24 |

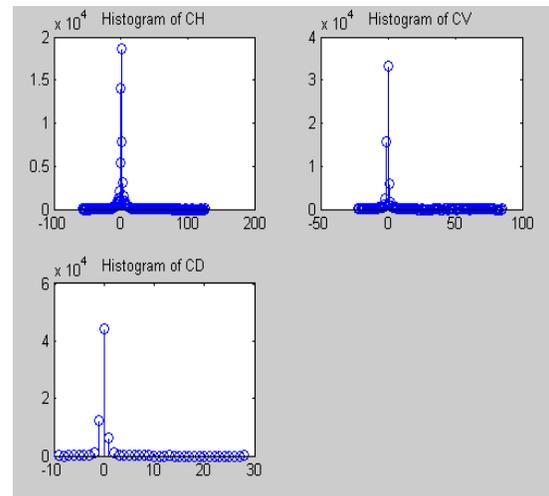

Figure 8 Histogram of Cameraman Image after IWT

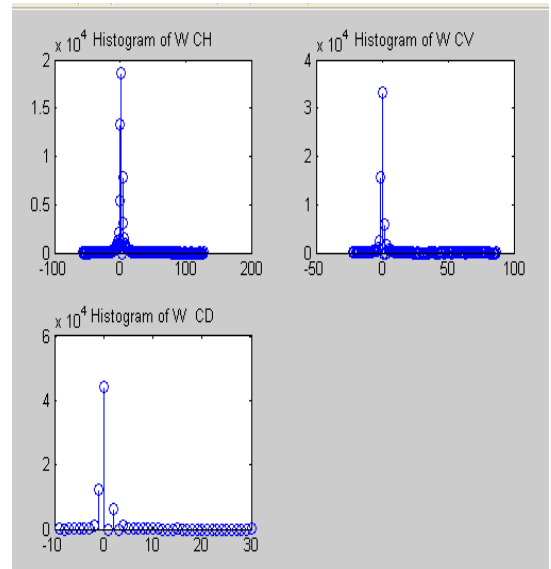

Figure 9 Histogram of Watermarked Cameraman Image

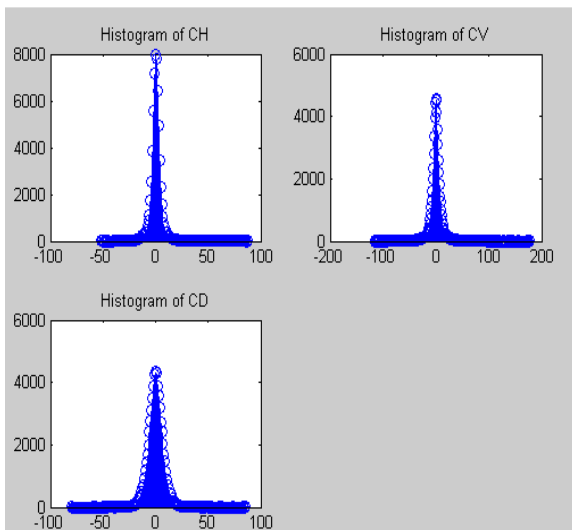

Figure 10 Histogram of Sailboat Image after IWT

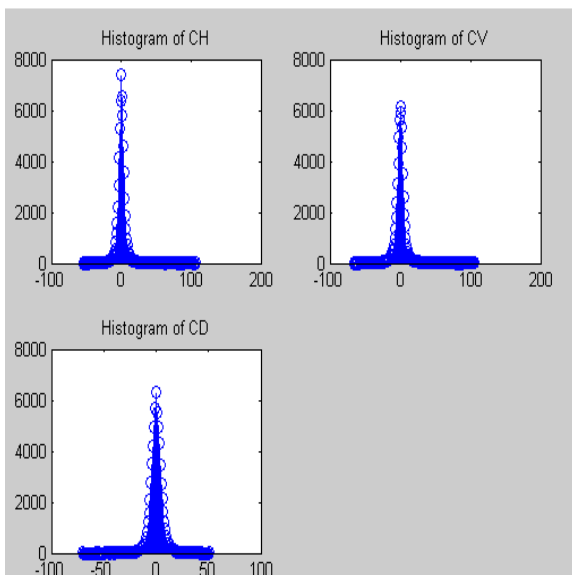

Figure 11 Histogram of Lena Image after IWT

Histogram of wavelet transformed cameraman image shows more number of coefficient values at peak point compared to Lena image and sailboat. This influences the embedding capacity. Cameraman image has higher embedding capacity compared to Lena or sailboat image. This is illustrated in figure 8, 10 and 11. Figure 9 shows the watermarked cameraman image wavelet histogram. This shows the shifted positions of the histogram points due to shifting and embedding.

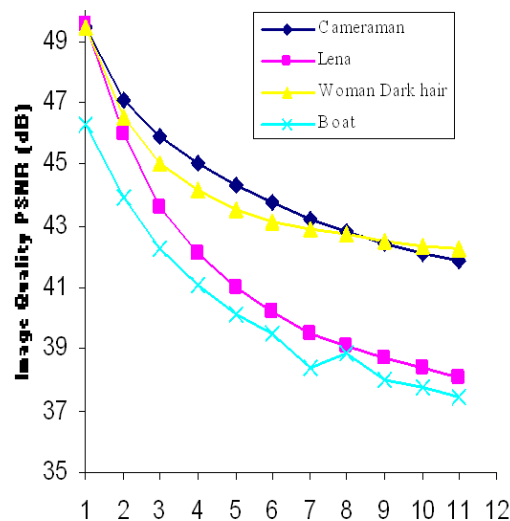

Figure 12 Image Quality Tested by using different number of embedding points in histogram

Image quality is tested using different number of embedding points in histogram to embed the watermark data. Each coefficient value can embed watermark bits equal to the number of occurrence of that point in the wavelet histogram. Figure 12 shows image quality decreases as we use more and more points in the histogram for embedding data. With lesser payload fewer points are used and we get more image quality for the watermarked images.

## VI. Conclusion

Reversible image watermarking using histogram shifting method was done and tested using different images. Embedding capacity not only varies from image to image, it also varies for various wavelets .The wavelet histogram is used foe embedding as it has a Laplacian like distribution and embedding can be done on both sides of the histogram to embed more data. More image quality is achieved for the same payload compared to other reversible watermarking methods this is a blind watermarking method. Original image and the embedded data are extracted exactly without any loss because our method is completely reversible. Images with more number of points on the wavelet histogram peak can embed more data.

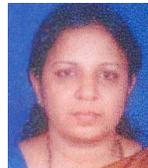

**S. Kurshid Jinna** Completed her B.E in Electronics and Communication Engineering from Thiagarajar College of Engineering, Madurai, in 1985 and M.E(Hons) in Computer Engineering from VJTI , University of Mumbai and doing Ph.D in faculty of information and communication in Anna University, Chennai. She is currently working as Professor & head of the department, Computer Science and Engineering in PET Engineering College, Vallioor, India.

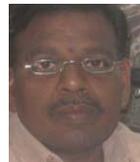

**Dr. L.Ganesan** completed his B.E in Electronics and Communication Engineering from Thiagarajar College of Engineering, Madurai and M.E in Computer Science and Engineering from Government College of Technology, Coimbatore. He completed his Ph.D from Indian Institute of Technology, Kharagpur in the area image processing. He has authored more than fifty publications in reputed International Journals. His area of interest includes image processing, multimedia and compressions. He is currently working as head of the department of Computer science and engineering, A.C. College of Engg. And Technology, Karaikudi, India